\documentclass[letterpaper,twocolumn,10pt]{article}
\usepackage{usenix2019_v3}

\usepackage{tikz}
\usepackage{amsmath}
\usepackage{xspace}

\newcommand{\myparagraph}[1]{\noindent\textbf{#1.}}

\begin{document}

\date{}

\title{\Large \bf Multi-version Indexing in Flash-based Key-Value Stores}

\author{
{\rm Pulkit A. Misra}\\
Duke University
\and
{\rm Jeffrey S. Chase}\\
Duke University
\and
{\rm Johannes Gehrke}\\
Microsoft Corporation
\and
{\rm Alvin R. Lebeck}\\
Duke University
} 

\newcommand{\kname}{\textsc{Semel}\xspace}
\newcommand{\tname}{\textsc{Milana}\xspace} 
\newcommand{\fname}{\textsc{SkimpyFTL}\xspace}

\maketitle

\begin{abstract}
Maintaining multiple versions of data is popular in key-value stores since it increases concurrency and improves performance. However, designing a multi-version key-value store entails several challenges, such as additional capacity for storing extra versions and an indexing mechanism for mapping versions of a key to their values. We present \fname, a FTL-integrated multi-version key-value store that exploits the remap-on-write property of flash-based SSDs for multi-versioning and provides a tradeoff between memory capacity and lookup latency for indexing.

\end{abstract}

\section{Introduction}
\label{sec:intro}

\begin{figure}[tb]
\centering
\includegraphics[height=1.7in,keepaspectratio]{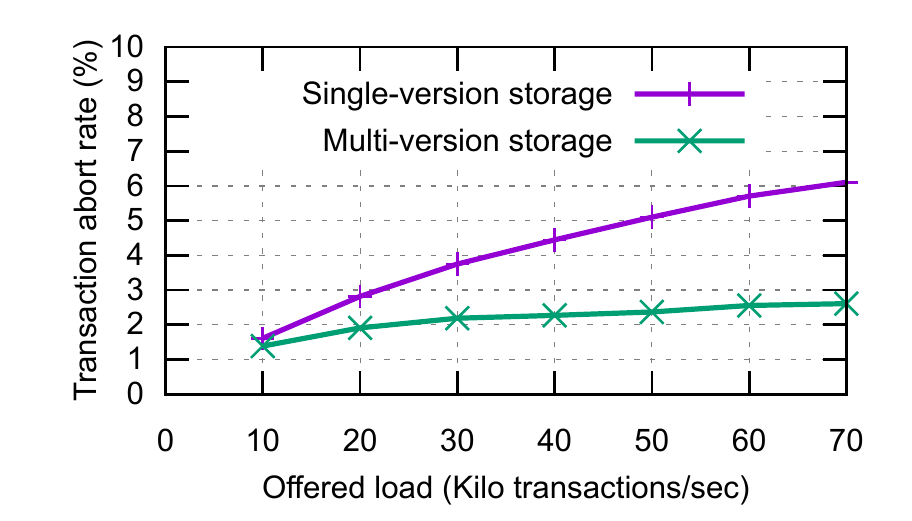}
\caption{Transaction abort rate with single vs multi-version storage}
\label{fig:svaborts}
\end{figure}

Transactional key-value stores use a multi-version storage to increase concurrency~\cite{corbett2013spanner} and reduce abort rate~\cite{bernstein1983multiversion} as reads can be satisfied from a consistent snapshot in the past (old versions), while writes create new versions. Figure~\ref{fig:svaborts} illustrates the impact of single vs multi-versioning on transaction abort rate for a social network application; workload and methodology are described in \S\ref{sec:eval}. As seen from the figure, multi-versioning reduces abort rate by 2$\times$ vs single-version storage, and its benefit increases with offered load.


However, there are several challenges in designing a multi-version storage, including: 1) additional capacity, 2) index for mapping versions to values, and 3) version management. First, the extra versions require additional capacity, which can be prohibitively expensive with in-memory (DRAM) storage systems. Second, these systems need an index to map versions of a key to their value so reads can be serviced from a consistent snapshot. A na\"ive approach is to store the entire index in DRAM; this approach provides the lowest lookup latencies but has a high space overhead. An efficient indexing technique needs to find a tradeoff between lookup latencies and DRAM requirement for indexing. Third, multi-versioning necessitates a version management (garbage collection) scheme for effective capacity utilization. The scheme needs to strike a balance between capacity reclamation by discarding old versions and servicing reads from a consistent snapshot.

Our previous work \kname~\cite{misra2017enabling} addresses the capacity and version management challenge with designing a multi-version storage system. It uses Solid State Drives (SSDs) for storing multiple versions of data and a watermark-based version management for effective capacity utilization as watermarks provide a bound on the oldest version that can be read by the application. This paper addresses the challenge with indexing.

SSDs are a more attractive proposition for multi-versioning than DRAM or newer non-volatile memory (NVM) technologies (e.g., Intel Optane) because they provide TB capacity per drive for less than \$1.00/GB. In addition, new standards for Software-Defined Flash (SDF) enable flash-based storage systems customized for application requirements~\cite{Josephson:2010:DFS,zhang2012deindirection,saxena2012flashtier,marmol2014nvmkv,huang2017flashblox,misra2017enabling}. These advances in flash storage enable storing more data per server, while delivering better performance compared to spinning disks and at a lower cost compared to DRAM and NVMs.

Furthermore, \kname exploits an intrinsic property of flash-based SSDs --- remap-on-write --- to implement multi-versioning in an SSDs Flash Translation Layer (FTL) using SDF. \kname's approach removes abstractions and provides better performance compared to a na\"ive approach of stacking a multi-version software layer over a standard FTL. However, \kname incurs a high space overhead since it maintains the entire index for mapping a version of a key to its location on flash in host DRAM.


This paper presents \fname, a system that addresses all the 3 challenges with designing a multi-version key-value storage system. It builds on top of \kname by leveraging SSDs for FTL-integrated multi-versioning along with a watermark-based version management scheme and addresses the challenge with indexing by providing a tradeoff between DRAM capacity and lookup latency. \fname uses a hash table for mapping key versions to their values. It follows SkimpyStash~\cite{debnath2011skimpy} in offloading hash collision list on flash to reduce DRAM requirement for indexing and adds support for multi-versioning using version pointers, which are also stored on flash along with the collision list.

A \fname prototype utilizing LightNVM Open-Channel SSD emulation framework~\cite{lightnvm2017} reveals \fname provides 72-91\% throughput of \kname for read-dominant key-value workloads (75-100\% reads), while reducing the memory requirement for indexing by a factor of 0.95$\times$. For a transactional YCSB~\cite{cooper2010ycsb} workload, \fname provides 85\% peak throughput of \kname. Finally, \fname outperforms a na\"ive multi-version key-value store implemented over a standard FTL on both workloads.

This paper is a first step in exploring the design space of memory-efficient indexing in multi-version flash-based key-value stores. Our overarching goal for this work is designing a distributed flash management framework. In the future, we plan to explore other data structures, such as LSM trees~\cite{oneil1996lsm}, for implementing multi-versioning inside the FTL. We also plan to further explore garbage collection strategies while maintaining multiple versions of data.


\section{Background}
\label{sec:background}
This section summarizes the internals and the remap-on-write property of a NAND-flash SSD (\S\ref{sec:internals}), provides background on Software-Defined Flash (SDF, \S\ref{sec:sdf}) and shows how \kname uses SDF in a multi-version storage system (\S\ref{sec:semel}).

\subsection{Internals of a NAND Flash-based SSD}
\label{sec:internals}
NAND flash memory in an SSD is organized as an array of blocks where each block
contains some number of pages. Typically pages are 2-16KB in size and each block contains 128-256 pages. The page size is the smallest unit for reads and writes. A page on flash must first be erased before it can be overwritten. However, erase operations on flash occur only at a block granularity.  


To accommodate these properties, flash-based SSDs use a Flash Translation Layer (FTL) to  map logical addresses dynamically to physical locations. This level of indirection allows the FTL to remap a logical block to a clean physical page on each write ({\em remap-on-write}), leaving the old value in place pending garbage collection.  The FTL's garbage collector also remaps values as needed to clean blocks to erase.   

This remap-on-write behavior naturally leaves previous versions of data in place for reading~\cite{subramanian2014snapshots}. \kname and \fname exploit this property as a foundation for an FTL-integrated multi-version key-value store. 


\subsection{Software-Defined Flash}
\label{sec:sdf}

The recently proposed Software-Defined Flash (SDF) structure moves some FTL functionality to host software~\cite{ouyang2011bbio,Josephson:2010:DFS,ouyang2014sdf,caulfield2012providing}. Several vendors provide some form of SDF (e.g., CNEX Labs, SanDisk/FusionIO, Radian Memory). 


SDF enables several optimizations across traditional system boundaries. First, flash-based key-value stores can map keys directly to pages on flash rather than addressing the SSD as a block device, eliminating one level of indirection~\cite{marmol2014nvmkv,zhang2012deindirection,Josephson:2010:DFS,saxena2012flashtier}. Next, customized mapping techniques may exploit system-specific information to improve performance and/or provide new functionality, such as snapshot capability~\cite{subramanian2014snapshots}. Another work leverages SDF to exploit the inherent parallelism of SSDs by mapping log-structured merge (LSM) tree~\cite{oneil1996lsm} operations  to different SSD channels~\cite{ouyang2014sdf}.

\subsection{\kname Flash Translation Layer}
\label{sec:semel}
\begin{figure}[t]
\centering
 \includegraphics[width=0.48\textwidth,keepaspectratio]{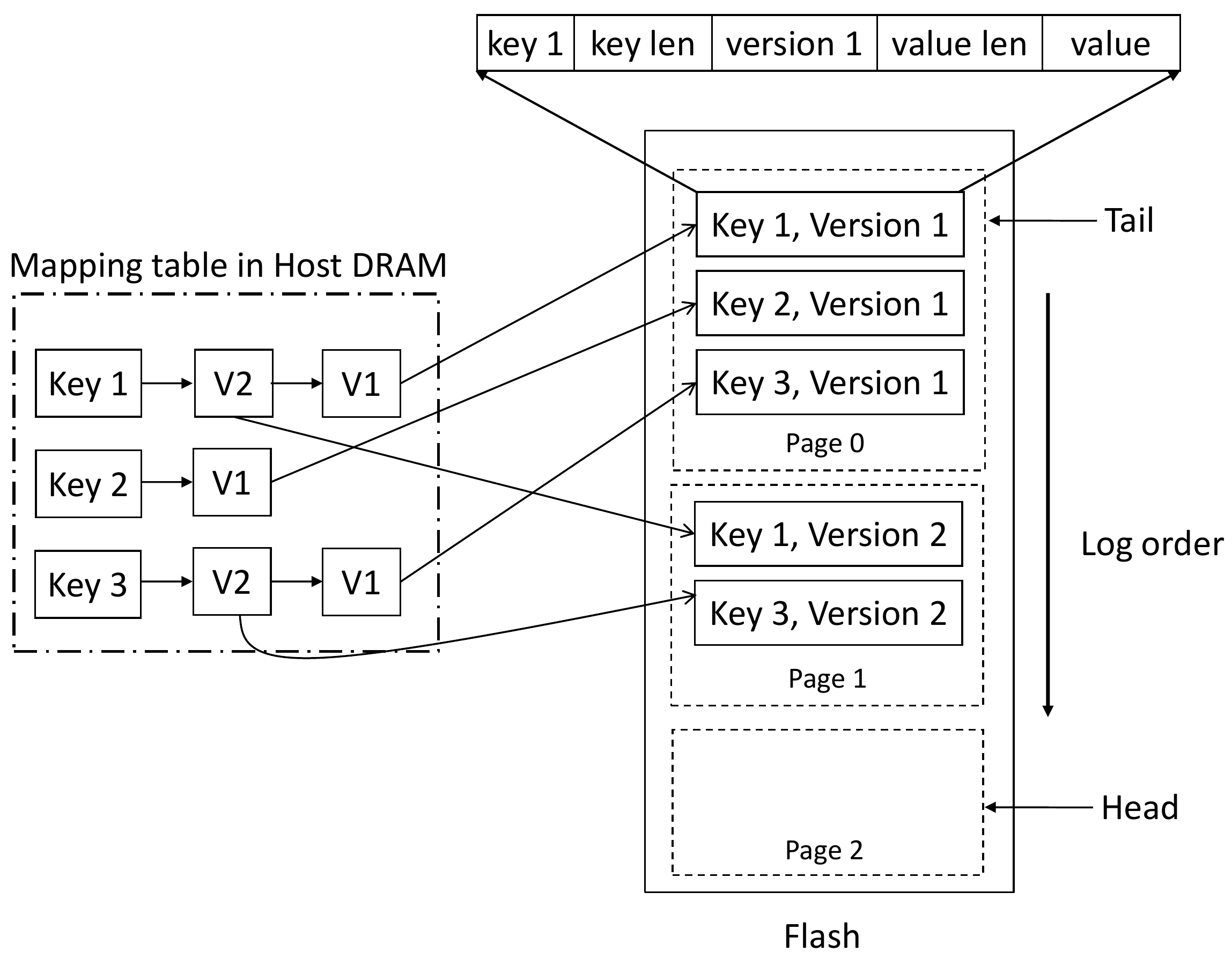}
\caption{Mapping table and data layout on flash in \kname}
\label{fig:semel}
\end{figure}

\kname is a lightweight multi-version key-value store based on SDF.   It writes new values in a log-structured fashion~\cite{rosenblum1991LFS}, densely packed in pages on flash.
Figure~\ref{fig:semel} shows the mapping table and data layout in \kname. As seen from the figure, \kname maintains a linked list in DRAM with an entry for each version of a key. Each version is assigned a 64-bit create timestamp and maps directly to a page on flash and the version's offset within the page, removing a level of indirection. 

\kname's DRAM-based mapping table is prohibitively expensive. Each version entry is 20B in size (4B page address, 8B version timestamp and 8B pointer to prior version): for a 1 TB SSD and 512B key-value pairs, \kname consumes 40 GB of DRAM to map the entire SSD.  The goal of \fname is to provide DRAM-efficient version mapping with performance as close as possible to \kname. 

\section{SkimpyFTL}

This section describes how \fname addresses the need for memory-efficient dynamic indexing of a multi-version store in \kname. 
Our approach combines a flash-based mapping table (\S\ref{sec:mapping_table}) with a DRAM-based mapping translation cache for hot keys (\S\ref{sec:key_cache}).  The two structures operate together to handle reads and writes efficiently in the common case (\S\ref{sec:req_life_cycle}).
The scheme also extends the garbage collection protocol (\S\ref{sec:gc}).

\subsection{Mapping Table}
\label{sec:mapping_table}
\fname indexes the key-value pairs on flash with a hash-based mapping table whose buckets are rooted in host DRAM.  Multiple key-value pairs may hash to the same bucket; \fname handles such collisions using {\em linear chaining}, where key-value pairs in the same bucket are chained using a linked list. Rather than maintaining this collision list in DRAM, \fname offloads it to flash; a hash table bucket in DRAM points to the head of the linear list on flash, and each entry in the list on flash points to the next entry. This approach is inspired by a prior work~\cite{debnath2011skimpy}. \fname allows configuring the number of hash buckets to achieve a desired balance of DRAM cost and lookup time.

\begin{figure}[t]
\centering
\includegraphics[width=0.5\textwidth,keepaspectratio]{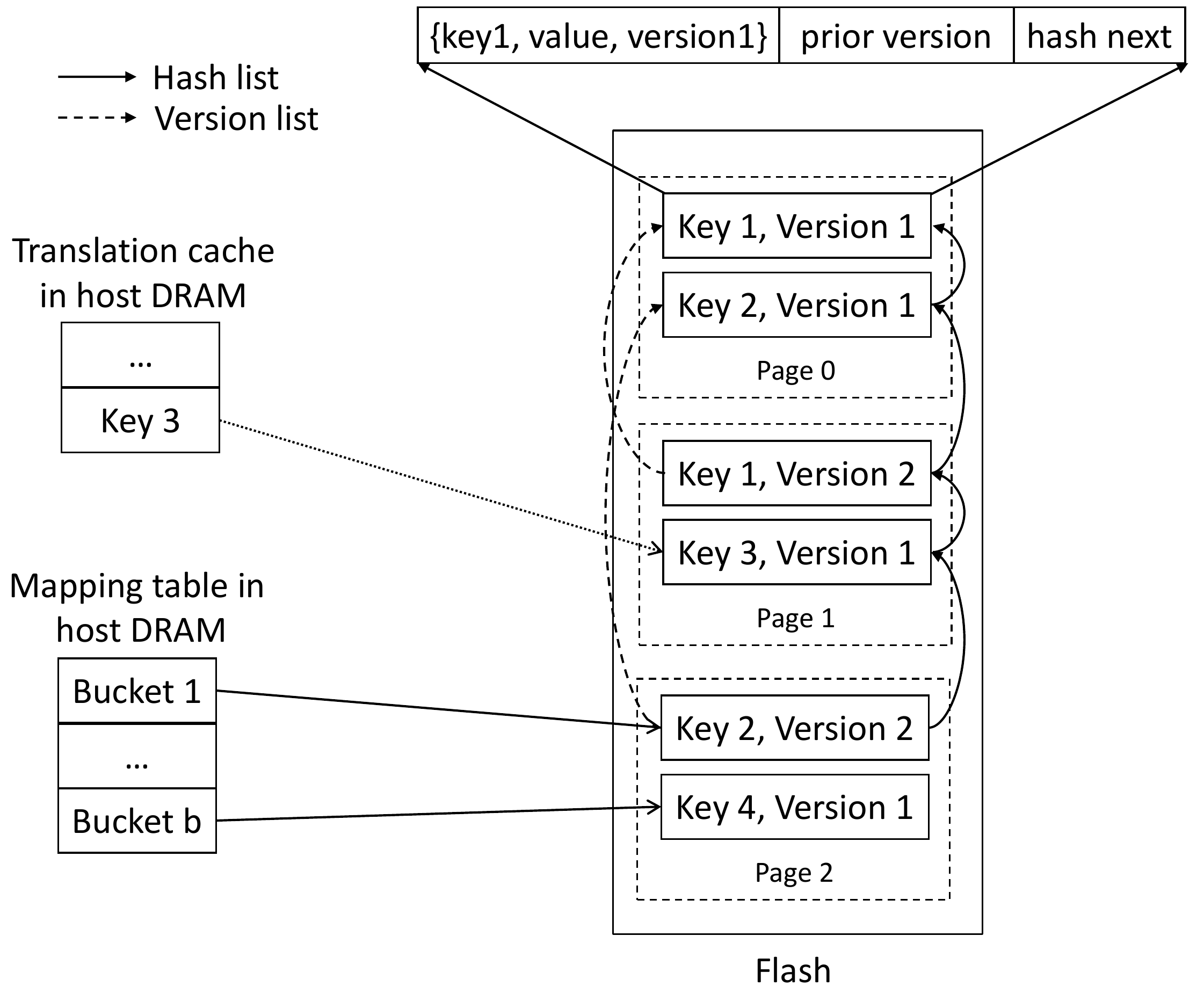}
\caption{Mapping table, translation cache and data layout on flash in \fname}
\label{fig:sftl}
\end{figure}

Figure~\ref{fig:sftl} shows the mapping table and the data layout on flash.  \fname writes data versions to flash in a log-structured fashion, as in \kname. In addition to the usual fields  for each key-value pair (version), \fname stores two pointers: a {\em hash next} pointer to the next entry in the bucket's collision list (hash chain) and a {\em prior version} pointer to a prior (older) version of the key.  The prior version pointer is an optimization: \fname can traverse a hash chain to find a requested version, but it is often faster to traverse the prior version pointer from a later version of the desired key, as described next.



\subsection{Mapping Translation Cache}
\label{sec:key_cache}
Prior characterization studies suggest that datacenter workloads tend to be read-dominated~\cite{atikoglu2012workload,nishtala2013memcached}. Furthermore, the popularity of data items in real-world workloads often follows a power law distribution, where a small subset of the keys receive a large portion of the accesses~\cite{cooper2010ycsb,atikoglu2012workload}. Such workloads would cause significant read amplification to traverse the collision lists.  In particular, a key that is updated infrequently tends to migrate to the end of the list, since new versions are written to the front of the list.

To mitigate this cost, \fname caches key translations in a {\it mapping translation cache} in DRAM.  A translation cache entry for a key stores the location of its latest version on flash.


\subsection{Request Life Cycle}
\label{sec:req_life_cycle}
Figure~\ref{fig:sftl} illustrates how the DRAM-based translation cache operates in conjunction with the mapping table to handle GET (read) and PUT (write) requests efficiently.

A GET request first does a lookup to locate the latest version of the requested key.  Then, if the request is a snapshot read for a previous version, it traverses the prior version pointer(s) to locate the requested version.  A lookup for a hot or recent key hits in the translation cache, which returns the pointer to the latest version.  On a cache miss, \fname hashes the key to a bucket and then follows the hash chain until it finds an entry for the key, which is the latest version.  It then caches the mapping in the translation cache.

For example, for a GET of \{key 1, version 1\} in Figure~\ref{fig:sftl}, \fname first hashes key 1 to bucket 1 and traverses the hash chain: \{key 2, version 2\} $\rightarrow$ \{key 3, version 1\} $\rightarrow$ \{key 1, version 2\}, until it encounters key 1. It then updates the  translation cache to point to the latest version of key 1.  

A PUT request hashes the key to a bucket and links the new value to the front of the hash chain: it sets its hash next field to the current index pointed to by the bucket and updates the current index.  To populate the prior version field, \fname probes the translation cache for the most recent version of the key.  This lookup typically results in a hit in the common case of a read-modify-write operation on the key.  On a miss, \fname sets the prior version field to a special value that indicates to a GET request that prior versions may exist and must be retrieved by searching further in the chain.   It fills in the missing prior version pointers during remapping (\S\ref{sec:gc}).

For example, for a PUT request for key 2 in Figure~\ref{fig:sftl}, \fname writes the new version to the end of the log on flash, hashes it to bucket 1, sets the hash next of key 2 to point to key 3 (on page 1) and updates bucket 1 to point to the new version.  It sets the prior version field of key 2 to point to a prior version.

\subsection{Garbage Collection}
\label{sec:gc}
The garbage collection process starts from the tail of the log and remaps valid versions (key-value pairs) to the head of the log.   For each key, \fname retains the youngest version with a timestamp less than the current {\it watermark}, and discards older versions.
The watermark is a timestamp that advances continuously.  In the \kname transactional key-value store, the watermark is the minimum timestamp that could appear in any future request from a client, following Centiman~\cite{ding2015centiman}.  Each client periodically passes its timestamp for its last acknowledged operation.  The minimum of these timestamps is the watermark.

\fname extends \kname's garbage collection to use the bucket collision lists and version lists, and to maintain their integrity.  To determine whether to remap or discard a version, \fname probes the translation cache and version list for a more recent version that is younger than the watermark.  On a miss, it must traverse a bucket hash chain.  For each remapped version, it must also update the hash next pointer of its predecessor in the chain to point at the new location.  For this reason, \fname garbage collects an entire hash bucket at a time by sweeping its hash chain and remapping retained versions in reverse temporal order, updating their pointers in the usual way as it goes.

To maintain fidelity of the hash chains and version lists, \fname blocks any new writes to a bucket during the process of traversing its hash chain and remapping valid data. The remapping process packs the retained versions of a bucket's keys densely into flash pages, reducing lookup time for later GET requests.
\section{Evaluation}
\label{sec:eval}
\begin{figure*}[tb]
	\centering
	\begin{minipage}[b]{0.33\linewidth}
	\centering
	\includegraphics[width=\textwidth]{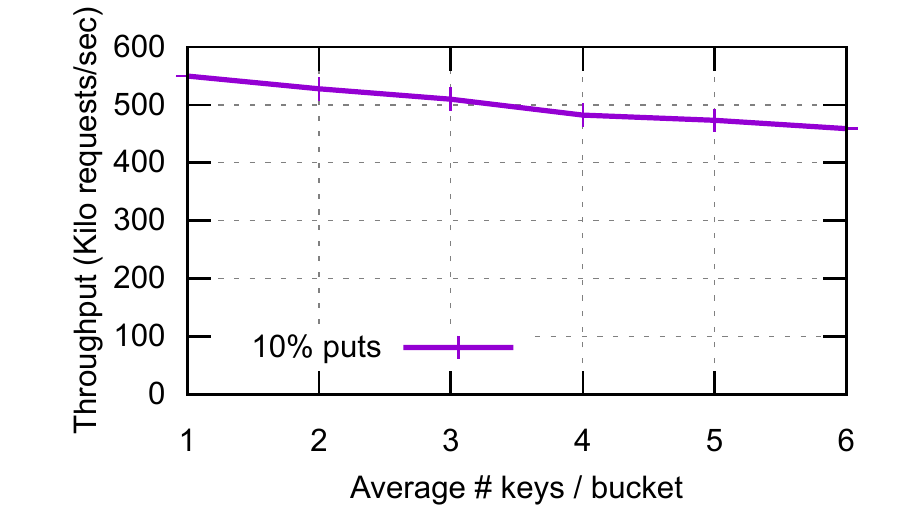}
	\caption{Impact of \# keys / bucket}
	\label{fig:buckets}
	\end{minipage}
	\begin{minipage}[b]{0.33\linewidth}
	\centering
	\includegraphics[width=\textwidth]{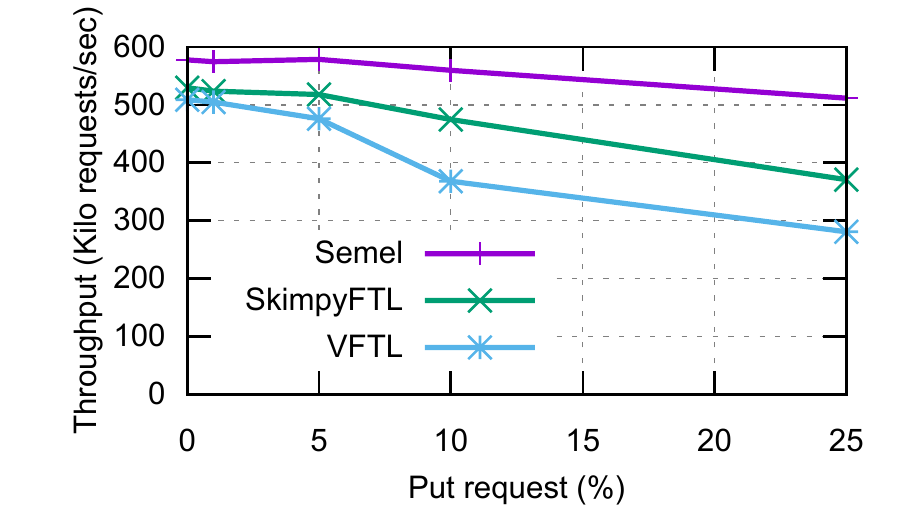}
	\caption{K-V operations throughput}
	\label{fig:kv_thr}
	\end{minipage}
    \begin{minipage}[b]{0.33\linewidth}
    	\centering
    	\includegraphics[width=\textwidth]{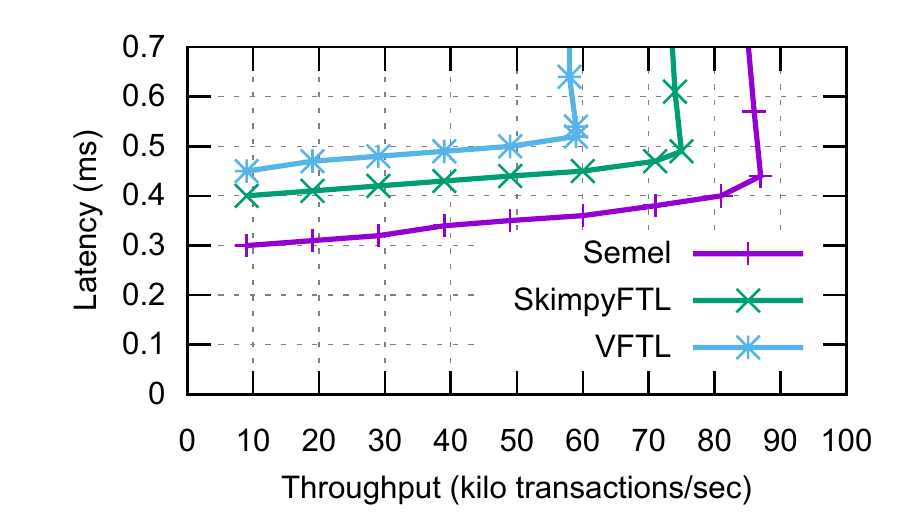}
    	\caption{Txns: throughput vs latency}
    	\label{fig:txn_thr_lat}
    \end{minipage}
\end{figure*}

Here we present the preliminary results for our prototype implementation of \fname and compare it with \kname. To elucidate the advantages of implementing multi-versioning within the FTL, we also compare against a separate multi-version KV store on top of a standard FTL; we refer to this system as VFTL. We use \kname with multi-versioning disabled for single-version store. 

\myparagraph{Implementation}
We use a modified Open-Channel SSD framework \cite{lightnvm2017} from our prior work~\cite{misra2017enabling} for all FTL implementations. In software-only mode, the emulator supports storing data values in DRAM, and provides IOCTLs for get, put and erase functionality for flash blocks. It also allows specifying latencies for read page, write page and block erase operations.

All our FTL implementations use 32 bits (4 bytes) for storing the location of a key on flash. To allow an FTL to pack multiple key-value pair within a page; we divide each page into fixed number of chunks and use 3 out of the 32 bits for storing the start chunk for a key in a page. The remaining 29 bits are used for addressing a page.

\myparagraph{Workload} We use two types of workloads for evaluation: 1) key-value operations workload, and 2) transactional YCSB~\cite{cooper2010ycsb} workload. We implement a micro-benchmark for the key-value operations workload; the micro-benchmark issues non-transactional get and put requests to single keys, for a varying get request percentage (\%). Popularity of keys in the benchmark is controlled using a zipfian distribution; we set the zipfian coefficient $\alpha$ to 0.99 in our micro-benchmark, a frequently used value in key-value workloads~\cite{cooper2010ycsb,atikoglu2012workload}. Our transactional YCSB~\cite{cooper2010ycsb} workload models a social network application, where the data of popular users is read more often and users have different rate of posting updates. The workload models this behavior by using different values of $\alpha$ for controlling popularity of keys in read-only ($\alpha_r$) and read-write ($\alpha_{rw}$) transactions. Each read-only transaction accesses from 1 to 10 keys, and a read-write transaction operates on 1 to 5.

\myparagraph{Experimental Setup} All experiments are run on a server with a 16 core Intel Xeon E5-2640 processor clocked and 128 GB DRAM. The SSD emulator is backed by 32 GB DRAM, with a hardware queue depth of 128. The SSD has a page size of 4KB and there are 32 pages in a block. A flash page read, write time is 50 and 100 $\mu s$ respectively and it takes 1 ms to erase a flash block. To ease garbage collection, all FTL implementations reserve 10\% capacity for remapping data.

For all experiments, we populate the system with 20M keys; the key size is 16B and packed data on flash is 512B, which includes key, value, version, and pointers to prior version and next entry in hash collision chain. As a flash page is 4KB in size, we employ a {\em packing logic} in the FTL that waits for up to 1 ms (tunable) to pack data of multiple keys into a page. There are up to 8 keys packed into a page, which increases the garbage collection overhead per page.  
Each run is of 15 mins and we pre-condition the SSD so garbage collection runs in the background in all experiments that perform writes.

Our \kname prototype stores the entire mapping table in DRAM. It needs $\sim$ 67M mappings (=32GB/512B) to address the entire SSD, and each mapping is 20B (4B page address, 8B version timestamp, and 8B prior version pointer) in size. Through sensitivity analysis, we set the translation cache size to 10\% of the number of keys in the workload, which gives $\sim$ 90\% hit rate for our current workload. Figure~\ref{fig:buckets} shows the impact of average number of keys / bucket on throughput with our key-value micro-benchmark with 10\% writes. Based on the results, we size the mapping table to have 5 keys / bucket i.e., 4M buckets. Each entry in the mapping table is 4B (page address), and 20B (4B page address, 16B LRU pointers) in the translation cache. \fname uses $\sim$ 5\% of the memory used by \kname.


\subsection{Key-Value Workload} 
We first evaluate the performance of all systems with a non-transactional key-value workload. 
Figure~\ref{fig:kv_thr} shows the throughput for varying put request percentage (\%) with our key-value micro-benchmark. The throughput of all 3 systems --- \kname, VFTL and \fname --- drops as put request percentage is increased because writes increase the garbage collection overhead. As expected, \kname provides the highest throughput as it implements multi-versioning in flash and maps all versions of all keys in DRAM. \fname provides from 72\% - 91\% of the throughput of \kname, while only using 5\% of the memory. The throughput degradation is worse in \fname as put request percentage increases because garbage collection overhead --- traversing entire hash collision chains for determining data to discard and remap --- is more frequent. We plan to explore other approaches to garbage collection, with lower overheads, in future. Interestingly, the throughput of VFTL is even lower than \fname. VFTL provides the lowest throughput as it suffers from log stacking~\cite{yang2014dont} --- the log in multi-version layer and standard FTL operate independently, which leads to uncoordinated garbage collection and randomization of writes.

\subsection{Transactional Workload}
To evaluate performance for a transactional workload, we stack a layer over the 3 systems that supports transactions using MVCC~\cite{bernstein1983multiversion}. The memory overhead of the layer is the same for all systems. Our YCSB~\cite{cooper2010ycsb} workload issues 90\% read-only transactions, with zipfian coefficient for read-only and read-write transactions set to $\alpha_r = 0.99$ and $\alpha_{rw} = 0.75$, respectively.

Figure~\ref{fig:txn_thr_lat} shows the throughput and latency with the 3 storage systems for an increasing offered load. The trend in performance of the systems is similar to the key-value workload. \kname provides the best performance; \fname provides 85\% of the peak throughput of \kname. VFTL provides the lowest peak throughput and highest latency. Thus, showing the disadvantage of the na\"ive approach of implementing multi-versioning, agnostic of the underlying storage medium. All approaches have similar transaction commit rates before saturation ($\sim$ 96\%).


\section{Related Work}
\label{sec:related}
Prior works have proposed indexes for flash-based key-value stores~\cite{andersen2009fawn, lim2011silt, marmol2014nvmkv,lu-wisckey,debnath2011skimpy,debnath2010flashstore,li2010fd}. However, none of these works support multi-versioning. FD-tree~\cite{li2010fd} and WiscKey~\cite{lu-wisckey} use a tree-based approach for indexing key-value pairs on SSD, whereas FAWN~\cite{andersen2009fawn}, NVMKV~\cite{marmol2014nvmkv} and FlashStore~\cite{debnath2010flashstore} use a hash table for indexing. SILT~\cite{lim2011silt} stores data in multiple tiers; as  key-value pairs age, they are compacted with other pairs and transitioned to other skimpy memory-optimized tiers. \fname is inspired from SkimpyStash \cite{debnath2011skimpy}, both approaches use configurable number of hash buckets to reduce the DRAM requirement for the mapping table, however SkimpyStash does not maintain multiple versions, nor does it have a translation cache.
\section{Conclusion}
We present \fname, a FTL-integrated multi-version key-value store that exploits remap-on-write property of flash-based SSDs for maintaining multiple versions of data and provides a tradeoff between DRAM capacity and lookup latency for indexing. Our evaluation reveals \fname provides 72-91\% throughput of \kname, while reducing memory requirement by a factor of 0.95x. We also show the benefit of implementing multi-versioning in the FTL. In the future, we plan to explore other data structures for indexing, garbage collection strategies and storing data in multiple tiers, based on versions, frequency and type of access to keys.
\section*{Acknowledgments}
This work is supported in part by the National Science Foundation (CNS-1616947).

\bibliographystyle{plain}
\bibliography{paper}

\end{document}